\documentclass[aps,pre,twocolumn,showpacs,preprintnumbers,amsmath,amssymb,floatfix]{revtex4-1}
\usepackage{graphicx}
\usepackage{hyperref}
\usepackage{amsfonts}
\usepackage{CJK}
\begin{document}
\bibliographystyle{apsrev4-1}
\hypersetup{hidelinks}
% Use the \preprint command to place your local institutional report
% number in the upper righthand corner of the title page in preprint mode.
% Multiple \preprint commands are allowed.
% Use the 'preprintnumbers' class option to override journal defaults
% to display numbers if necessary
%\preprint{}

%Title of paper
\title{Equilibrium Equality for Free Energy Difference}

\author{Weitao Chen}\email{phycwt@stu.xmu.edu.cn}

\affiliation{Department of Physics, Key Laboratory of Low Dimensional Condensed Matter Physics (Department of Education of Fujian Province), and Jiujiang Research Institute, Xiamen University, Xiamen 361005, Fujian, China }

\date{\today}
\begin{abstract}
    Jarzynski Equality (JE) and the thermodynamic integration method are conventional methods to calculate free energy difference (FED) between two equilibrium states with constant temperature of a system. However, a number of ensemble samples should be generated to reach high accuracy for a system with large size, which consumes a lot computational resource. Previous work had tried to replace the non-equilibrium quantities with equilibrium quantities in JE by introducing a virtual integrable system and it had promoted the efficiency in calculating FED between different equilibrium states with constant temperature. To overcome the downside that the FED for two equilibrium states with different temperature can't be calculated efficiently in previous work, this article derives out the Equilibrium Equality for FED between any two different equilibrium states by deriving out the equality for FED between states with different temperatures and then combining the equality for FED between states with different volumes. The equality presented in this article expresses FED between any two equilibrium states as an ensemble average in one equilibrium state, which enable the FED between any two equilibrium states can be determined by generating only one canonical ensemble and thus the samples needed are dramatically less and the efficiency is promoted a lot. Plus, the effectiveness and efficiency of the equality are examined in Toda-Lattice model with different dimensions.
\end{abstract}

% insert suggested PACS numbers in braces on next line
\pacs{PACS}
% insert suggested keywords - APS authors don't need to do this
%\keywords{}
%\maketitle must follow title, authors, abstract, \pacs, and \keywords
\maketitle
    \section{Introduction}\label{sec1}
    Free energy is a characteristic function of a thermodynamic system. According to the second law of thermodynamics, free energy difference (FED) represents the maximum work that a system can output in an isothermal process. There are broad applications \cite{ref1} of FED calculation especially in interdisciplinary like biophysics and chemical physics.

    However, the numerical calculation of FED is a tricky problem. Till now, theoretical physicists have worked out some numerical methods to calculate FED. The thermodynamic integration method \cite{ref2} is a conventional method to calculate FED, which is based on a quasi-equilibrium process to do the integration. To make this method practical, a path that connect initial state and terminal state should be designed and the path should be divided into many sub-paths as tiny as possible to ensure the thermodynamic process remain quasi-equilibrium. In the sub-paths, the equilibrium quantities of the system including internal energy and pressure should be calculated because the integration should be worked out based on these quantities. If $F,T,V$ represents the free energy, temperature and volume of a thermodynamic system, there are thermodynamic relations
    \begin{equation}\label{eq1}
        {{\left[ \frac{\partial (F/T)}{\partial (1/T)} \right]}_{V}}=U,
    \end{equation}
    and
    \begin{equation}\label{eq2}
       {{\left( \frac{\partial F}{\partial V} \right)}_{T}}=-p,
    \end{equation}
    where $p$ and $U$ are pressure and internal energy of the system. The FED between different states can be determined by integrating these quantities. To reach high accuracy, a large enough ensemble is needed to calculate the average of these quantities in every step. Therefore, the total samples needed for this method is numerous, which makes the cost of computation is demanding.

    In 1997, Jarzynski derived out the Jarzynski Equality (JE) \cite{ref3,ref4}, which also be named as Non-equilibrium Equality for Free Energy Difference
    \begin{equation}\label{eq3}
        {{e}^{-\beta \Delta F}}={{\left\langle {{e}^{-\beta w}} \right\rangle }_{a}}.
    \end{equation}
    Here $\beta \equiv 1/\left( {{k}_{B}}T \right)$ is the reverse temperature of the system, $\Delta F\equiv {{F}_{b}}-{{F}_{a}}$ is FED between state $a$ and state $b$ , $w$ is the work done to the system by the environment when the system evolves from state $a$ to state $b$. The angular bracket and the subscript $a$ at the right hand side (r.h.s.) of Eq.(\ref{eq3}) represents the average over the canonical ensemble of state $a$. JE represents FED between state $a$ and state $b$ as the work done to the system when the system evolves from the initial state to the terminal state and the system needn't to be in equilibrium when it approaches state $b$. Besides, the system also needn't to be coupled with the heat reservoir to remain constant temperature all the time, it only should be ensured that the state $a$ and state $b$ remain in the same temperature \cite{ref3,ref4,ref5}. Compared to the thermodynamic integration method, although methods based on JE needn't to calculate pressure and internal energy of intermediate states, it demands a number of samples to reach high accuracy because it contains the calculation of the work done in a non-equilibrium process. Especially for a system with large size or a relatively larger change of the volume, it is even tough to calculate FED with JE. Therefore, many works have been done to overcome the default of JE, a through survey can be found in Ref. \cite{ref6} . However, although much effort had paid to deal with the problem, the algorithms based on JE are still less efficient than conventional thermodynamic integration method or the thermodynamic perturbation method. To overcome these downside, later work \cite{ref7} gave out an equilibrium equality for FED between states with different volume but constant temperature and the efficiency of the equality is superior to the thermodynamic integration method. In fact, the equality is a generalization of free energy perturbation theory (FEPT) \cite{ref8}, which fixes the difficulty that the FEPT is unable to calculate FED between two equilibrium states with different volume. Nevertheless, although the equality can calculate FED between states with different volume efficiently, it is unable to calculate FED between states with different temperature. Hence, this article tries to derive out an equilibrium equality which is capable to calculate FED between any different equilibrium states.

    \section{Equilibrium Equality for FED between states with different volume}\label{sec2}
    To achieve an equilibrium equality which is capable to calculate FED between any different states, this article will introduce an intermediate state whose volume and parameters are the same as initial state and temperature is the same as the terminal state. By this way, the calculation of FED can be done by combining the FED between states with different volume and the FED between states with different temperature. Here the FED between states with constant temperature but different volume can be determined by the equilibrium equality for FED between states with constant temperature but different volume \cite{ref7}. The Hamilton of a system consist of $N$ particles
    \begin{equation}\label{eq4}
      H=\sum\limits_{i=1}^{N}{\frac{\mathbf{p}_{i}^{2}}{2{{m}_{i}}}}+U(\mathbf{r}),
    \end{equation}
    where $\mathbf{r}\equiv ({{\mathbf{r}}_{1}},\cdots ,{{\mathbf{r}}_{N}})$ are the position vectors of particles, ${{\mathbf{p}}_{i}}$ is the momentum of the $i$-th particle and ${{m}_{i}}$ is the mass of the $i$-th particle. This work rewrote JE as an equality with only the average of the potential energy by introducing an integrable adiabatic piston model. The equality for FED between states with different volume can be expressed as
    \begin{equation}\label{eq5}
       {{e}^{-\beta \Delta F}}={{r}^{N}}{{\left\langle {{e}^{\beta \left[ U(\mathbf{x};{{L}_{a}})-U(r\mathbf{x};{{L}_{b}}) \right]}} \right\rangle }_{a,\mathbf{x}}},
    \end{equation}
    here $r\equiv {{L}_{b}}/{{L}_{a}}$, ${{L}_{a}}$ and ${{L}_{b}}$ are the corresponded volume of state $a$ and state $b$ which remain in constant temperature. And $\mathbf{x}\equiv \left( {{\mathbf{x}}_{1}},\cdots ,{{\mathbf{x}}_{N}} \right)$ are the coordinates of particles and the angular bracket at the r.h.s.of Eq.(\ref{eq5}) represents the average over the distribution ${{P}_{a}}(\mathbf{x})={{e}^{-{{\beta }_{a}}U(\mathbf{x})}}/\int{{{e}^{-{{\beta }_{a}}U(\mathbf{x})}}d\mathbf{x}}$. Note that the equality is not limited to one dimension, for system with multiple dimensions
    \begin{equation}\label{eq6}
     {{e}^{-\beta \Delta F}}={{\left( {{V}_{b}}/{{V}_{a}} \right)}^{N}}{{\left\langle {{e}^{\beta \left[ {{U}_{a}}-{{U}_{b}} \right]}} \right\rangle }_{a,\mathbf{r}}},
    \end{equation}
    here ${{V}_{a}},{{V}_{b}}$ are the volume of the system in state $a$ and state $b$, ${{U}_{a}}\equiv U(\mathbf{x},\mathbf{y},\mathbf{z};{{L}_{a,x}},{{L}_{a,y}},{{L}_{a,z}})$, ${{U}_{b}}\equiv U({{r}_{x}}\mathbf{x},{{r}_{y}}\mathbf{y},{{r}_{z}}\mathbf{z};{{L}_{a,x}},{{L}_{a,y}},{{L}_{a,z}})$, and the angular bracket at the r.h.s. of Eq.(\ref{eq6}) represents the average over the distribution ${{P}_{a}}(\mathbf{r})={{e}^{-{{\beta }_{a}}{{U}_{a}}}}/\int{{{e}^{-{{\beta }_{a}}{{U}_{a}}}}d\mathbf{r}}$.

    Consider more general case that a system with a set of other parameters named $\Gamma $, combine the FEPT \cite{ref8} and then the equality can be written as
    \begin{equation}\label{eq7}
     {{e}^{-\beta \Delta F}}={{r}^{N}}{{\left\langle {{e}^{\beta \left[ U(\mathbf{x};{{\Gamma }_{a}},{{L}_{a}})-U(r\mathbf{x};{{\Gamma }_{b}},{{L}_{b}}) \right]}} \right\rangle }_{a,\mathbf{x}}},
    \end{equation}
    equalities for multiple dimensions can be expressed in similar forms. The FED between states with constant temperature but different volume and parameters can be swiftly calculated by working out the average over only one canonical ensemble.

    Theoretically, Eq.(\ref{eq7}) is a novel equality to calculate FED between states with the same temperature but different volume and parameters. Numerically, the equality comes out to express FED as the average over only one canonical ensemble, which means that the calculation can be achieved by using numerical methods such as Monte Carlo algorithms directly to generate a distribution with the coordinates of particles as the variables. This advantage is conductive to reduce the cost of computational resource and promote the efficiency.

    \section{Equilibrium Equality for FED between states with different temperature}\label{sec3}
    The equilibrium equality for FED between states with the same volume but different temperature can be expressed in similar form as the case for constant temperature mentioned above. Consider a system evolves from initial state $a$ with temperature ${{T}_{a}}$ to terminal state $b$ with temperature ${{T}_{b}}$ and the Hamilton of the system is the same as Eq.(\ref{eq4}), according to the definitions, the free energy ${{F}_{a}}$ and the partial function ${{Z}_{a}}$ of the system in temperature ${{T}_{a}}$ can be presented as
    \begin{equation}\label{eq8}
      {{F}_{a}}=-\ln {{Z}_{a}}/{{\beta }_{a}},
    \end{equation}
    \begin{equation}\label{eq9}
    \begin{split}
      {{Z}_{a}}&=\underset{i}{\mathop{\Pi }}\,\int{{{e}^{-{{\beta }_{a}}\frac{\mathbf{p}_{i}^{2}}{2{{m}_{i}}}}}d{{\mathbf{p}}_{i}}}\cdot \int{{{e}^{-{{\beta }_{a}}U(\mathbf{r})}}d\mathbf{r}}\\
      &={{({{m}_{1}}\cdots {{m}_{N}})}^{d/2}}{{(2\pi )}^{Nd/2}}\beta _{a}^{-Nd/2}\cdot \int{{{e}^{-{{\beta }_{a}}U(\mathbf{r})}}d\mathbf{r}},
    \end{split}
    \end{equation}
    here $d$ is the dimension of the system. The free energy ${{F}_{b}}$ and partial function ${{Z}_{b}}$ of the system in state $b$ with temperature ${{T}_{b}}$ can be expressed in a similar way. According to Eq.(\ref{eq8}),
    \begin{equation}\label{eq10}
      {{e}^{-{{\beta }_{a}}{{F}_{a}}}}={{Z}_{a}}.
    \end{equation}

    Similarly,
    \begin{equation}\label{eq11}
     {{e}^{-{{\beta }_{b}}{{F}_{b}}}}={{Z}_{b}}.
    \end{equation}

    Define
    \begin{equation}\label{eq12}
     \zeta =\frac{{{T}_{a}}}{{{T}_{b}}}=\frac{{{\beta }_{b}}}{{{\beta }_{a}}},
    \end{equation}
    divide Eq. (\ref{eq10}) by Eq.(\ref{eq11}) and combine the corresponded partial function, the equality can be expressed as
    \begin{equation}\label{eq13}
    \begin{split}
    {{e}^{-{{\beta }_{a}}(\zeta {{F}_{b}}-{{F}_{a}})}}&=\frac{\beta _{b}^{-Nd/2}\cdot \int{{{e}^{-{{\beta }_{b}}U(\mathbf{r})}}d\mathbf{r}}}{\beta _{a}^{-Nd/2}\cdot \int{{{e}^{-{{\beta }_{a}}U(\mathbf{r})}}d\mathbf{r}}}\\
    &={{\zeta }^{-Nd/2}}\cdot \frac{\int{{{e}^{{{\beta }_{a}}[U(\mathbf{r})-\zeta U(\mathbf{r})]}}{{e}^{-{{\beta }_{a}}U(\mathbf{r})}}d\mathbf{r}}}{\int{{{e}^{-{{\beta }_{a}}U(\mathbf{r})}}d\mathbf{r}}}\\
    &={{\zeta }^{-Nd/2}}\cdot {{\left\langle {{e}^{(1-\zeta ){{\beta }_{a}}U(\mathbf{r})}} \right\rangle }_{a,\mathbf{r}}},
    \end{split}
    \end{equation}
    where the angular bracket at the r.h.s.of Eq.(13) represents the average over the distribution ${{P}_{a}}(\mathbf{r})={{e}^{-{{\beta }_{a}}U(\mathbf{r})}}/\int{{{e}^{-{{\beta }_{a}}U(\mathbf{r})}}d\mathbf{r}}$.

    Similarly, Eq.(\ref{eq13}) expresses FED between equilibrium states with different temperature as the average of a canonical ensemble, which means that the calculation of FED between states with different temperature but the same volume can be achieved by calculating the average over the canonical ensemble only once. Note that the equality can be also derived out by the FEPT \cite{ref8}. Consider a system with Hamilton
    \begin{equation}\label{eq14}
     \tilde{H}=\sum\limits_{i=1}^{N}{\frac{\mathbf{p}_{i}^{2}}{2{{m}_{i}}}}+\sigma U(\mathbf{r}),
    \end{equation}
    assume that the system evolves from initial equilibrium states with temperature ${{T}_{a}}$ and parameter $\sigma =1$ to terminal state with temperature ${{T}_{a}}$ and parameter $\sigma =\zeta$, the initial state is the equilibrium state with temperature ${{T}_{a}}$. If ${{F}_{a}}$ represents the free energy of the initial state and ${{\tilde{F}}_{a}}$ represents the free energy of the terminal state, according to the FEPT \cite{ref8},
    \begin{equation}\label{eq15}
    {{e}^{-{{\beta }_{a}}({{{\tilde{F}}}_{a}}-{{F}_{a}})}}={{\left\langle {{e}^{(1-\zeta ){{\beta }_{a}}U(\mathbf{r})}} \right\rangle }_{a,\mathbf{r}}},
    \end{equation}
    and according to the definition of free energy
    \begin{small}
    \begin{equation}\label{eq16}
    \begin{split}
    {{\tilde{F}}_{a}}&=-\ln {{\tilde{Z}}_{a}}/{{\beta }_{a}}\\
    &=-\ln \{{{({{m}_{1}}\cdots {{m}_{N}})}^{d/2}}{{(2\pi )}^{Nd/2}}\beta _{a}^{-Nd/2}\cdot \int{{{e}^{-{{\beta }_{a}}\zeta U(\mathbf{r})}}d\mathbf{r}}\}/{{\beta }_{a}}\\
    &=-\zeta (Nd/2)\ln \zeta /{{\beta }_{b}}-\zeta \ln {{Z}_{b}}/{{\beta }_{b}}\\
    &=-\zeta (Nd/2)\ln \zeta /{{\beta }_{b}}+\zeta {{F}_{b}},
    \end{split}
    \end{equation}
    \end{small}
    note that the Eq.(\ref{eq12}) is used to simplify the calculation of Eq.(\ref{eq16}). Eq.(\ref{eq13}) can be derived out by combing the Eq.(\ref{eq16}) and Eq.(\ref{eq15}), therefore, the equality is also a generalization of the FEPT for states with different temperature.
    \section{General Equality for FED}\label{sec4}
    For two states with different volume, temperature and parameters, the equilibrium equality for FED can be expressed in a more general form. If the Hamilton of a system
    \begin{equation}\label{eq17}
     H=\sum\limits_{i=1}^{N}{\frac{\mathbf{p}_{i}^{2}}{2{{m}_{i}}}}+U(\mathbf{x},\mathbf{y},\mathbf{z};{{L}_{x}},{{L}_{y}},{{L}_{z}},\Gamma ),
    \end{equation}
    where $\mathbf{x}\equiv ({{x}_{1}},\cdots ,{{x}_{N}})$, $\mathbf{y}\equiv ({{y}_{1}},\cdots ,{{y}_{N}})$, $\mathbf{z}\equiv ({{z}_{1}},\cdots ,{{z}_{N}})$, $({{x}_{i}},{{y}_{i}},{{z}_{i}})$ is the coordinates of the $i$-th particle, ${{L}_{x}}$, ${{L}_{y}}$, ${{L}_{z}}$ are lengths of the system in three dimensions and $\Gamma$ is the set of other parameters of the system.

    Assumed that the initial equilibrium state of the system named state $a$ can be expressed by the set $({{T}_{a}},{{L}_{x,a}},{{L}_{y,a}},{{L}_{z,a}},{{\Gamma }_{a}})$ including the temperature, volume and other parameters of the system, the terminal state named state $b$ can be also expressed by the set $({{T}_{b}},{{L}_{x,b}},{{L}_{y,b}},{{L}_{z,b}},{{\Gamma }_{b}})$. And the intermediate state named state $\tilde{b}$ whose volume and other parameters is the same as state $a$ and the temperature is the same as state $b$ state can be introduced with the set $({{T}_{b}},{{L}_{x,a}},{{L}_{y,a}},{{L}_{z,a}},{{\Gamma }_{a}})$. If ${{F}_{a}}$, ${{F}_{b}}$ and ${{F}_{{\tilde{b}}}}$ represent the free energy of state $a$, $b$ and $\tilde{b}$, and thus ${{F}_{a}}\equiv F({{T}_{a}},{{L}_{x,a}},{{L}_{y,a}},{{L}_{z,a}},{{\Gamma }_{a}})$,${{F}_{b}}\equiv F({{T}_{b}},{{L}_{x,b}},{{L}_{y,b}},{{L}_{z,b}},{{\Gamma }_{b}})$ and ${{F}_{{\tilde{b}}}}\equiv F({{T}_{b}},{{L}_{x,a}},{{L}_{y,a}},{{L}_{z,a}},{{\Gamma }_{a}})$.

    According to Eq.(\ref{eq6}) and Eq.(\ref{eq7}), the FED between state $b$ and state $\tilde{b}$ can be expressed as
    \begin{equation}\label{eq18}
     {{e}^{-{{\beta }_{b}}({{F}_{b}}-{{F}_{{\tilde{b}}}})}}=\frac{V_{b}^{N}}{V_{a}^{N}}\cdot \frac{\int{{{e}^{-{{\beta }_{b}}U({{r}_{x}}\mathbf{x},{{r}_{y}}\mathbf{y},{{r}_{z}}\mathbf{z};{{L}_{x,b}},{{L}_{y,b}},{{L}_{z,b}},{{\Gamma }_{b}})}}d\mathbf{x}d\mathbf{y}d\mathbf{z}}}{\int{{{e}^{-{{\beta }_{b}}U(\mathbf{x},\mathbf{y},\mathbf{z};{{L}_{x,a}},{{L}_{y,a}},{{L}_{z,a}},{{\Gamma }_{a}})}}d\mathbf{x}d\mathbf{y}d\mathbf{z}}},
    \end{equation}
    where ${{V}_{a}}={{L}_{x,a}}{{L}_{y,a}}{{L}_{z,a}}$, ${{V}_{b}}={{L}_{x,b}}{{L}_{y,b}}{{L}_{z,b}}$ and ${{r}_{x}}={{L}_{x,b}}/{{L}_{x,a}}$, ${{r}_{y}}={{L}_{y,b}}/{{L}_{y,a}}$, ${{r}_{z}}={{L}_{z,b}}/{{L}_{z,a}}$.

    According to Eq.(\ref{eq13}), the FED between state $\tilde{b}$ and state $a$ for $d=3$ can be expressed as
    \begin{equation}\label{eq19}
     {{e}^{-{{\beta }_{a}}(\zeta {{F}_{{\tilde{b}}}}-{{F}_{a}})}}=\frac{\beta _{b}^{-3N/2}}{\beta _{a}^{-3N/2}}\cdot \frac{\int{{{e}^{-{{\beta }_{b}}U(\mathbf{x},\mathbf{y},\mathbf{z};{{L}_{x,a}},{{L}_{y,a}},{{L}_{z,a}},{{\Gamma }_{a}})}}d\mathbf{x}d\mathbf{y}d\mathbf{z}}}{\int{{{e}^{-{{\beta }_{a}}U(\mathbf{x},\mathbf{y},\mathbf{z};{{L}_{x,a}},{{L}_{y,a}},{{L}_{z,a}},{{\Gamma }_{a}})}}d\mathbf{x}d\mathbf{y}d\mathbf{z}}},
    \end{equation}
    by combining Eq.(\ref{eq18}) and Eq.(\ref{eq19}), the FED between state $b$ and state $a$ can be expressed as
    \begin{equation}\label{eq20}
    \begin{split}
     &{{e}^{{{\beta }_{a}}{{F}_{a}}-{{\beta }_{b}}{{F}_{b}}}}\\&=\frac{V_{b}^{N}}{V_{a}^{N}}\cdot \frac{\beta _{b}^{-3N/2}}{\beta _{a}^{-3N/2}}\frac{\int{{{e}^{-{{\beta }_{b}}U({{r}_{x}}\mathbf{x},{{r}_{y}}\mathbf{y},{{r}_{z}}\mathbf{z};{{L}_{x,b}},{{L}_{y,b}},{{L}_{z,b}},{{\Gamma }_{b}})}}d\mathbf{x}d\mathbf{y}d\mathbf{z}}}{\int{{{e}^{-{{\beta }_{a}}U(\mathbf{x},\mathbf{y},\mathbf{z};{{L}_{x,a}},{{L}_{y,a}},{{L}_{z,a}},{{\Gamma }_{a}})}}d\mathbf{x}d\mathbf{y}d\mathbf{z}}}\\
     &=\frac{V_{b}^{N}}{V_{a}^{N}}\cdot \frac{\beta _{b}^{-3N/2}}{\beta _{a}^{-3N/2}}{{\left\langle {{e}^{{{\beta }_{a}}{{U}_{a}}-{{\beta }_{b}}{{U}_{b}}}} \right\rangle }_{a,\mathbf{r}}},
    \end{split}
    \end{equation}
    here ${{U}_{a}}\equiv U(\mathbf{x},\mathbf{y},\mathbf{z};{{L}_{x,a}},{{L}_{y,a}},{{L}_{z,a}},{{\Gamma }_{a}})$, ${{U}_{b}}\equiv U({{r}_{x}}\mathbf{x},{{r}_{y}}\mathbf{y},{{r}_{z}}\mathbf{z};{{L}_{x,b}},{{L}_{y,b}},{{L}_{z,b}},{{\Gamma }_{b}})$, the angular brackets at the r.h.s. of Eq.(\ref{eq20}) represent the average over the distribution
    \begin{equation*}
     {{P}_{a,\mathbf{r}}}(\mathbf{x},\mathbf{y},\mathbf{z})=\frac{{{e}^{-{{\beta }_{a}}U(\mathbf{x},\mathbf{y},\mathbf{z};{{L}_{x,a}},{{L}_{y,a}},{{L}_{z,a}},{{\Gamma }_{a}})}}}{\int{{{e}^{-{{\beta }_{a}}U(\mathbf{x},\mathbf{y},\mathbf{z};{{L}_{x,a}},{{L}_{y,a}},{{L}_{z,a}},{{\Gamma }_{a}})}}d\mathbf{x}d\mathbf{y}d\mathbf{z}}}.
    \end{equation*}

    Similarly, Eq.(\ref{eq20}) can be generalized to 1-D and 2-D cases. For the 2-D system with ${{V}_{a}}\equiv {{L}_{x,a}}{{L}_{y,a}}$ and ${{V}_{b}}\equiv {{L}_{x,b}}{{L}_{y,b}}$,
    \begin{equation}\label{eq21}
     {{e}^{{{\beta }_{a}}{{F}_{a}}-{{\beta }_{b}}{{F}_{b}}}}=\frac{V_{b}^{N}}{V_{a}^{N}}\cdot \frac{\beta _{b}^{-N}}{\beta _{a}^{-N}}{{\left\langle {{e}^{{{\beta }_{a}}{{U}_{a}}-{{\beta }_{b}}{{U}_{b}}}} \right\rangle }_{a,\mathbf{r}}},
    \end{equation}
    here ${{U}_{a}}\equiv U(\mathbf{x},\mathbf{y};{{L}_{x,a}},{{L}_{y,a}},{{\Gamma }_{a}})$, ${{U}_{b}}\equiv U({{r}_{x}}\mathbf{x},{{r}_{y}}\mathbf{y};{{L}_{x,b}},{{L}_{y,b}},{{\Gamma }_{b}})$, the angular bracket at the r.h.s. of Eq.(\ref{eq21}) represents the average over the distribution
    \begin{equation*}
     {{P}_{a,\mathbf{r}}}(\mathbf{x},\mathbf{y})=\frac{{{e}^{-{{\beta }_{a}}U(\mathbf{x},\mathbf{y};{{L}_{x,a}},{{L}_{y,a}},{{\Gamma }_{a}})}}}{\int{{{e}^{-{{\beta }_{a}}U(\mathbf{x},\mathbf{y};{{L}_{x,a}},{{L}_{y,a}},{{\Gamma }_{a}})}}d\mathbf{x}d\mathbf{y}}}.
    \end{equation*}

    For 1-D system,
    \begin{equation}\label{eq22}
     {{e}^{{{\beta }_{a}}{{F}_{a}}-{{\beta }_{b}}{{F}_{b}}}}=\frac{L_{x,b}^{N}}{L_{x,a}^{N}}\cdot \frac{\beta _{b}^{-N/2}}{\beta _{a}^{-N/2}}{{\left\langle {{e}^{{{\beta }_{a}}{{U}_{a}}-{{\beta }_{b}}{{U}_{b}}}} \right\rangle }_{a,\mathbf{x}}},
    \end{equation}
    here ${{U}_{a}}\equiv U(\mathbf{x};{{L}_{x,a}},{{\Gamma }_{a}})$, ${{U}_{b}}\equiv U({{r}_{x}}\mathbf{x};{{L}_{x,b}},{{\Gamma }_{b}})$, the angular bracket at the r.h.s. of Eq.(\ref{eq22}) represents the average over the distribution
    \begin{equation*}
     {{P}_{a,\mathbf{x}}}(\mathbf{x})=\frac{{{e}^{-{{\beta }_{a}}U(\mathbf{x};{{L}_{x,a}},{{\Gamma }_{a}})}}}{\int{{{e}^{-{{\beta }_{a}}U(\mathbf{x};{{L}_{x,a}},{{\Gamma }_{a}})}}d\mathbf{x}}}.
    \end{equation*}

    Eq.(\ref{eq20}), Eq.(\ref{eq21}) and Eq.(\ref{eq22}) have represented FED between states with different temperature, volume and parameters as the average over one canonical ensemble successfully. Theoretically, theses equalities complete the further generalization of FEPT and propose a brand-new method for the calculation of FED between any states. Numerically, the distribution of these equality only contains the coordinates of particles, which means that the distribution can be generated by the conventional methods to generate a canonical ensemble including Monte Carlo algorithm conveniently. What's more, the FED between any two states can be determined by calculating the average over only one canonical ensemble. The dramatic promotion of the efficiency to calculate FED can be foreseen because the kinetic energy is derived out analytically. Meanwhile, the samples needed to generate is far less than the conventional thermodynamic integration method when they reach the same accuracy, which make the calculation of FED between any states more time-saving and convenient.
    \section{FED for Toda-lattice Model}\label{sec5}
    To test the effectiveness and efficiency of Eq.(\ref{eq20}), Eq.(\ref{eq21}) and Eq.(\ref{eq22}), we apply the equalities to the Toda-lattice model with different dimensions. The potential energy of 1-D Toda-lattice model \cite{ref9}
    \begin{equation}\label{eq23}
     U=\sum{[{{e}^{-({{x}_{i+1}}-{{x}_{i}}-1)}}+(}{{x}_{i+1}}-{{x}_{i}}-1)],
    \end{equation}
    the two kinds of particles with mass 1 and 2 align alternately. Note that it is an non-integrable model \cite{ref10} with complexity of calculation.

\begin{figure}[t]
    \centering
    \includegraphics[width=1\linewidth]{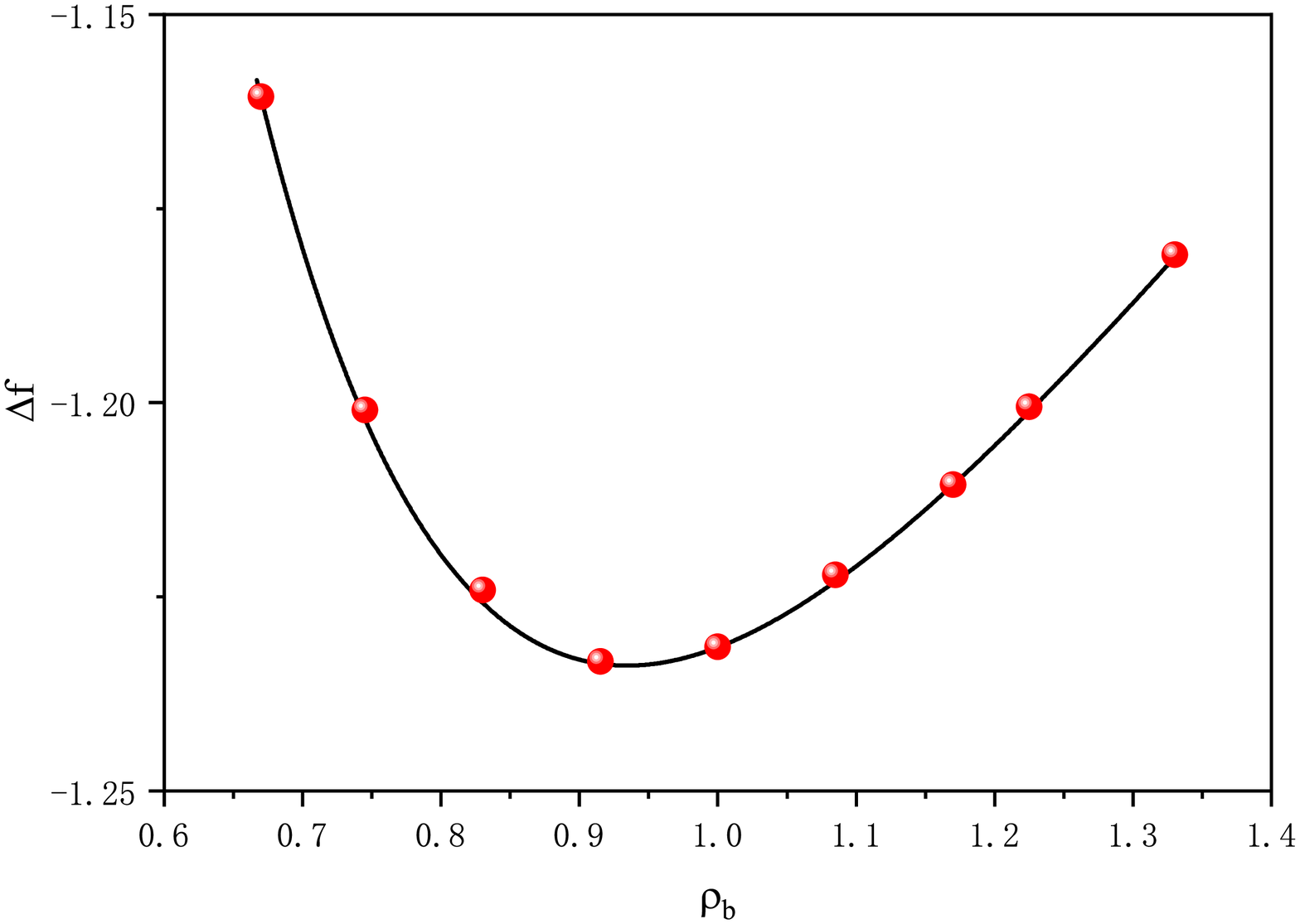}
    \caption{\label{fig1} FED per particle changes with the particle density of terminal state ${{\rho }_{b}}$. Here $N=20$ and the particle density of the initial state ${{\rho }_{a}}=\frac{2}{3}$, the reverse temperature of initial state ${{\beta }_{a}}=40$ and the reverse temperature of the terminal state ${{\beta }_{b}}=50$. The dots are results of Eq.(\ref{eq22}) with ensemble size $10$ and the solid line represents the results of thermodynamic integration method with ensemble size ${{10}^{9}}$.}
    \end{figure}

\begin{figure}[b]
    \centering
    \includegraphics[width=1\linewidth]{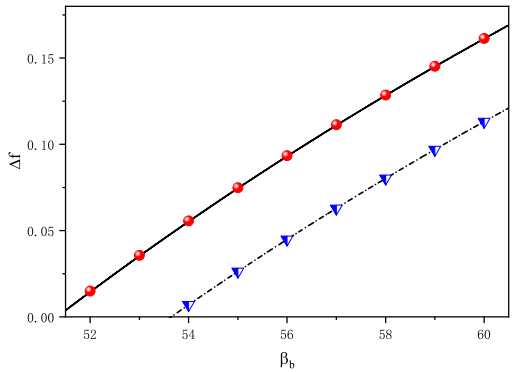}
    \caption{\label{fig2} FED per particle changes with the reverse temperature of terminal state ${{\beta }_{b}}$. Here $N=20$ and the particle density of the initial state ${{\rho }_{a}}=\frac{2}{3}$ and reverse temperature of initial state ${{\beta }_{a}}=50$. The dots and triangles are results of Eq.(\ref{eq22}) for particle density in terminal state ${{\rho }_{b}}=\frac{4}{3}({{L}_{b}}=15)$ and ${{\rho }_{b}}=1({{L}_{b}}=20)$ with ensemble size $10$, respectively, the solid line and the dash line represent the results of thermodynamic integration method with ensemble size ${{10}^{9}}$.}
    \end{figure}

    To examine the effectiveness and efficiency of Eq.(\ref{eq22}) in calculating the FED between states with different volume and temperature, we calculate the FED by applying Eq.(\ref{eq22} to the 1-D Toda-lattice model and compare its results with the results of the thermodynamic integration method. The fixed boundary condition is taken in the calculation, which means that the length between the two boundaries of the system remains $L$. Define the particle density $\rho \equiv N/L$, FED $\Delta F\equiv {{F}_{b}}-{{F}_{a}}$ and corresponded FED per particle $\Delta f\equiv \Delta F/N$, Fig.(\ref{fig1}) and Fig.(\ref{fig2}) present the computational results of Eq.(\ref{eq22}) and thermodynamic integration method. Here we set ${{F}_{a}}=0$, the Fig.(\ref{fig1}) shows the curve of FED changing with the particle density of terminal state when the temperature of the initial state and terminal state is fixed and the Fig.(\ref{fig2}) shows the curve of FED changing with the reverse temperature of terminal state when the volume of the initial state and terminal state is fixed. The results of Eq.(\ref{eq22}) and thermodynamic integration method compared in the figures meet perfectly, which confirms the effectiveness of Eq.(\ref{eq22}). Note that all the statistical uncertainty ("error bar") of data points in the figures is too small to be presented in the figure. The canonical ensemble Monte Carlo algorithm is used to generate the microscopic states of the initial state, the ensemble size is $10$ and all the corresponded uncertainty of $\Delta f$ is smaller than ${{10}^{-7}}$. Similarly, the intermediate states of the thermodynamic integration method are generated by the canonical ensemble Monte Carlo algorithm, the ensemble size is ${{10}^{9}}$and the corresponded uncertainty of $\Delta f$ is smaller than ${{10}^{-5}}$. In comparison, the ensemble sizes needed of Eq.(\ref{eq22}) is far less than the thermodynamic integration method and thus the computational time is about $4\times {{10}^{-7}}$ of the thermodynamic integration method, which suggests that the equality is efficient enough.

\begin{figure}[t]
    \centering
    \includegraphics[width=1\linewidth]{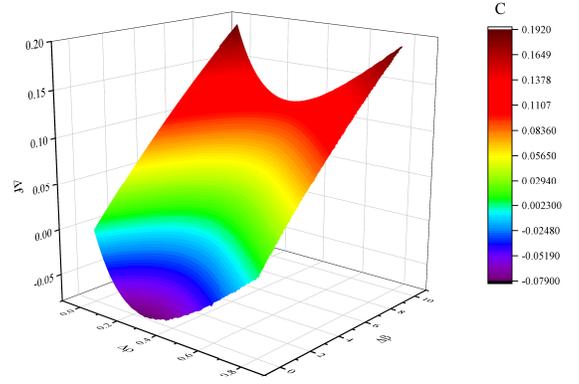}
    \caption{\label{fig3}  FED per particle changes with the particle density difference $\Delta \rho $ and the reverse temperature difference $\Delta \beta $. The particle density of the initial state ${{\rho }_{a}}=\frac{2}{3}$,the reverse temperature of initial state ${{\beta }_{a}}=50$ and the average ensemble size is $10$ for every data point.}
    \end{figure}

\begin{figure}[b]
    \centering
    \includegraphics[width=1\linewidth]{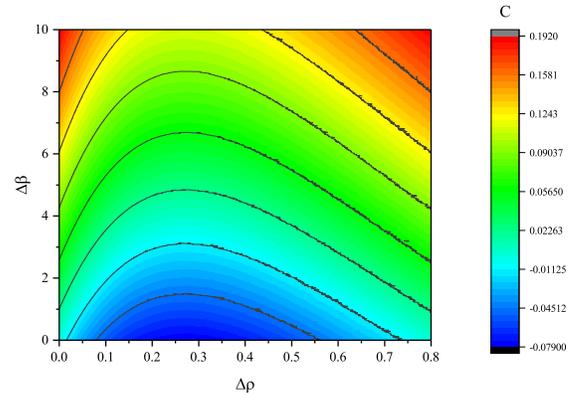}
    \caption{\label{fig4}  The corresponded contour map of FIG.3.}
    \end{figure}

     Meanwhile, take the advantage of time-saving and high efficiency of Eq.(\ref{eq22}), we compute the FED per particle $\Delta f$ changes with the particle density difference $\Delta \rho $ and the reverse temperature difference $\Delta \beta $. Fig.(\ref{fig3}) shows the distribution of $\Delta f$ in 3-D and Fig.(\ref{fig4}) is the corresponded contour map. Similarly, the canonical ensemble Monte Carlo algorithm is used in the computation, the ensemble size is $10$ and the corresponded uncertainty of $\Delta f$ is smaller than ${{10}^{-7}}$. The free energy landscape like Fig.(\ref{fig3}) or Fig.(\ref{fig4}) is useful and essential to study the properties of thermodynamic system in the fields of Biophysics or Chemical thermodynamics and thus the equality enable the computation to be achieved in a smaller cost.

    The multiple dimensional Toda-lattice Model can be studied by applying Eq.(\ref{eq20}) and Eq.(\ref{eq21}). Consider a square lattice with ${{N}_{x}}\times {{N}_{y}}$ sites or a cubic lattice with ${{N}_{x}}\times {{N}_{y}}\times {{N}_{z}}$ sites, the potential energy can be expressed as
    \begin{equation}\label{eq24}
    U=\sum{\left[ {{e}^{-(|{{r}_{i}}-{{r}_{j}}|-1)}}+(|{{r}_{i}}-{{r}_{j}}|-1) \right]},
    \end{equation}
    where the sum runs over both $i$ and $j$ satisfying that the $i$-th and the $j$-th particles are the nearest neighbors and meanwhile $i<j$.

\begin{figure}[h]
    \centering
    \includegraphics[width=1\linewidth]{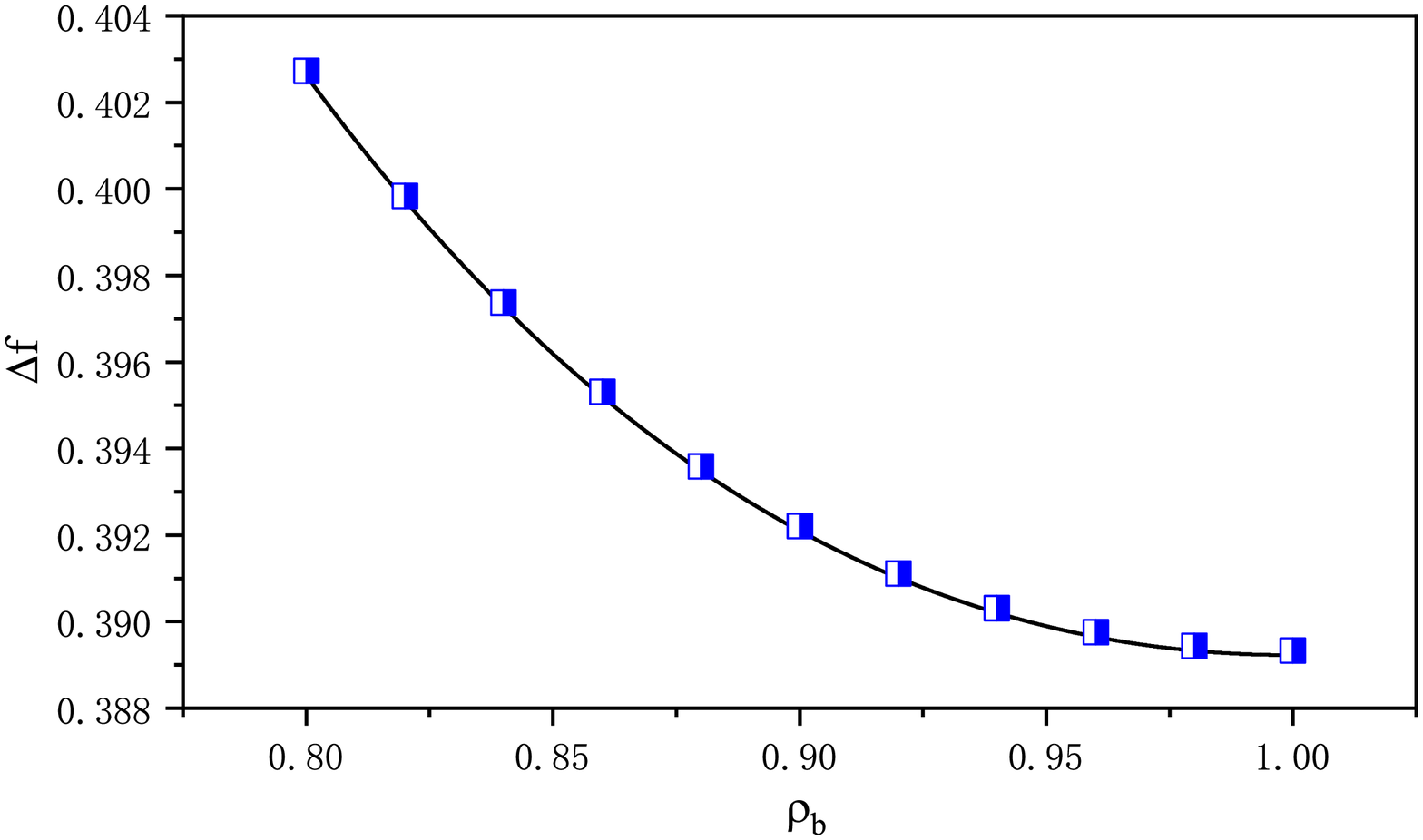}
    \caption{\label{fig5}   FED per particle changes with the particle density of terminal state ${{\rho }_{b}}$ for 2-D case. Here ${{N}_{x}}\times {{N}_{y}}=8\times 8$ and the particle density of the initial state ${{\rho }_{a}}=0.8$, the reverse temperature of initial state ${{\beta }_{a}}=8\times {{10}^{3}}$ and the reverse temperature of the terminal state ${{\beta }_{b}}={{10}^{4}}$. The squares are results of Eq.(\ref{eq21}) with ensemble size $10$ and the solid line represents the results of thermodynamic integration method with ensemble size ${{10}^{9}}$.}
    \end{figure}

\begin{figure}[h]
    \centering
    \includegraphics[width=1\linewidth]{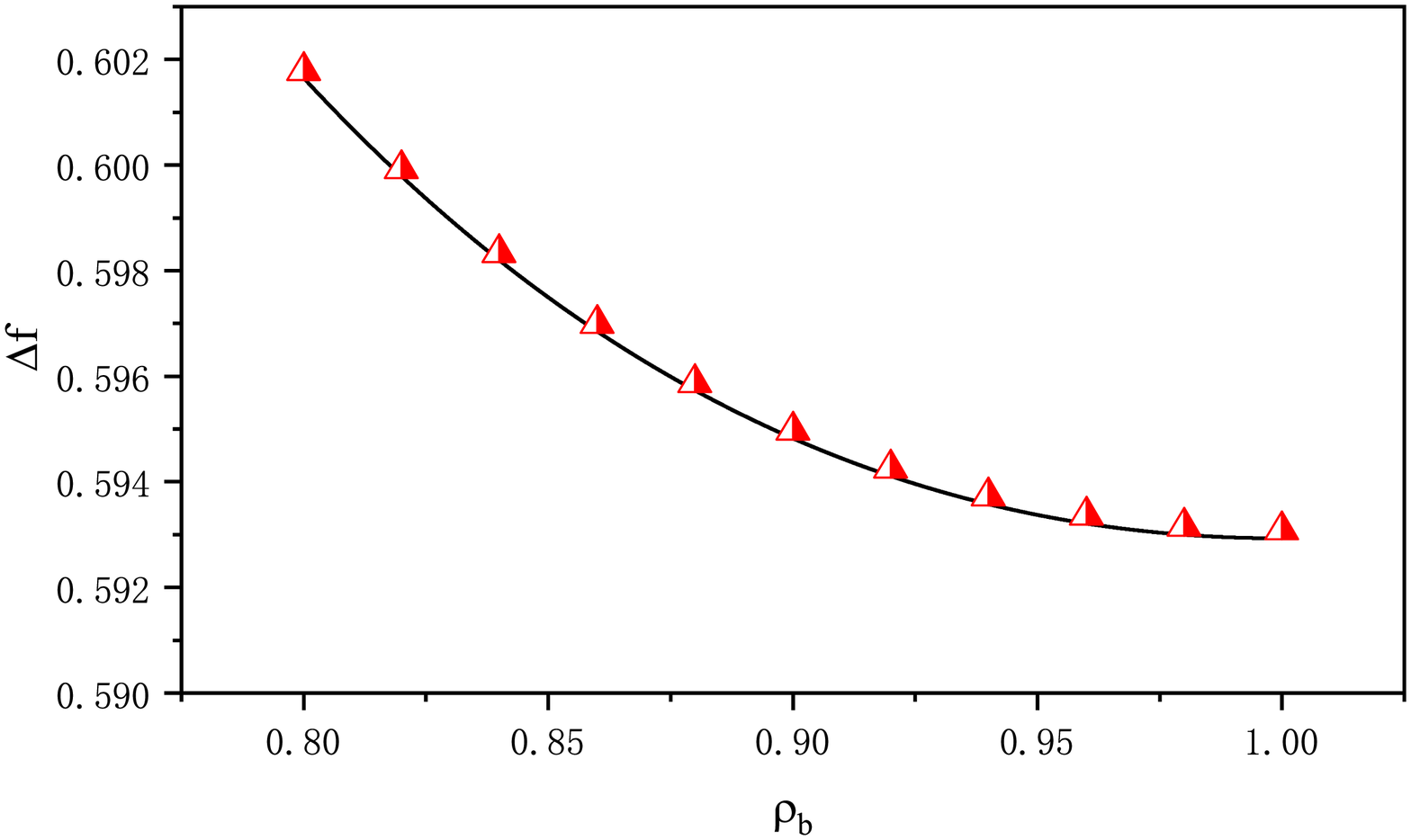}
    \caption{\label{fig6} FED per particle changes with the particle density of terminal state ${{\rho }_{b}}$ for 3-D case. Here ${{N}_{x}}\times {{N}_{y}}\times {{N}_{z}}=8\times 8\times 8$ and the particle density of the initial state ${{\rho }_{a}}=0.8$ , the reverse temperature of initial state ${{\beta }_{a}}=8\times {{10}^{3}}$ and the reverse temperature of the terminal state ${{\beta }_{b}}={{10}^{4}}$. The triangles are results of Eq.(\ref{eq20}) with ensemble size $10$ and the solid line represents the results of thermodynamic integration method with ensemble size ${{10}^{9}}$.}
    \end{figure}

\begin{figure}[h]
    \centering
    \includegraphics[width=1\linewidth]{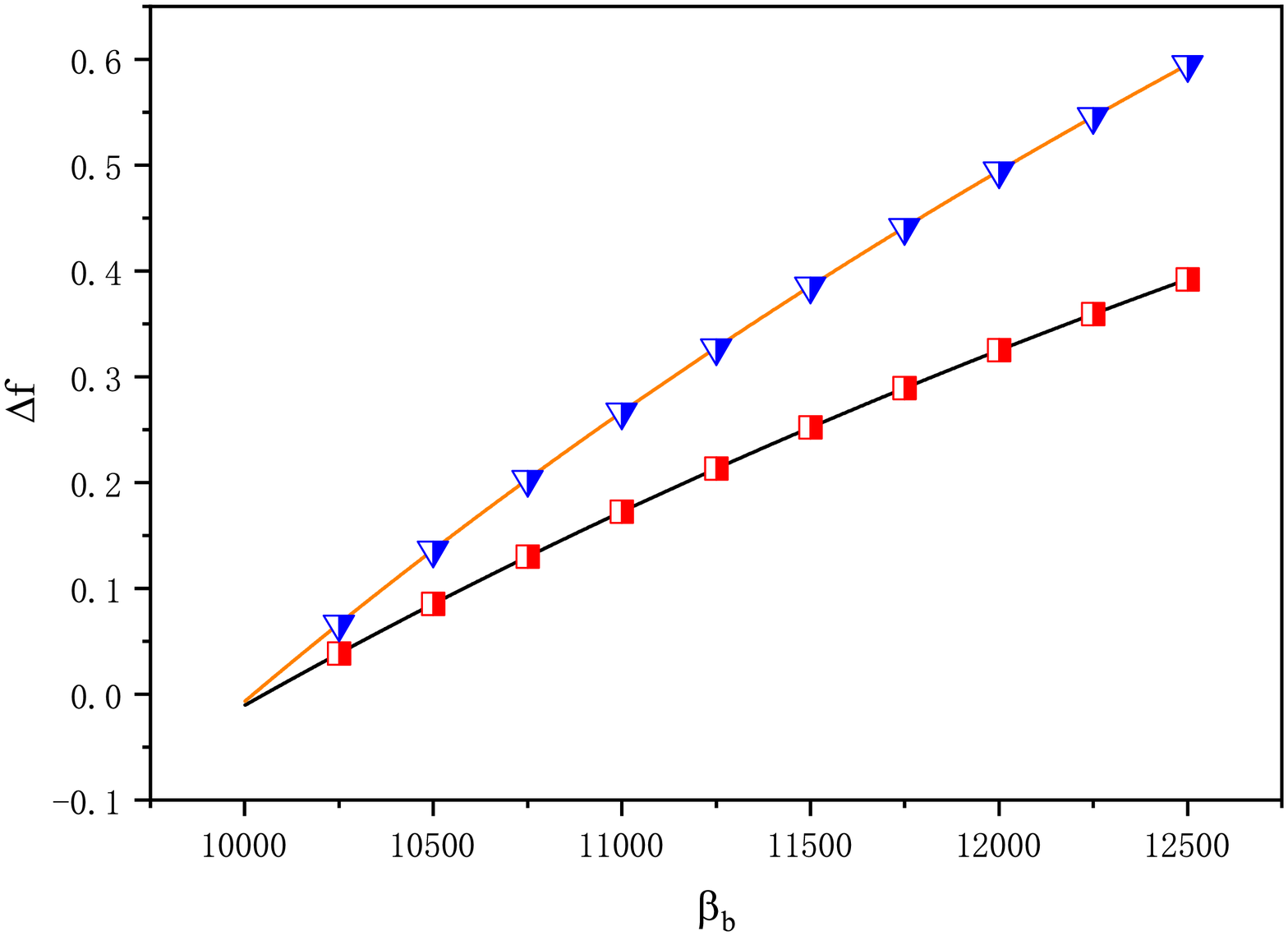}
    \caption{\label{fig7}   FED per particle changes with the reverse temperature of terminal state ${{\beta }_{b}}$ for 2-D and 3-D case. Here ${{N}_{x}}\times {{N}_{y}}=8\times 8$ for a square lattice and ${{N}_{x}}\times {{N}_{y}}\times {{N}_{z}}=8\times 8\times 8$ for a cubic lattice. The particle density of the initial state ${{\rho }_{a}}=0.8$, reverse temperature of initial state ${{\beta }_{a}}={{10}^{4}}$ and the particle density of the terminal state ${{\rho }_{b}}=1.0$ for both cases. The squares and triangles are results of Eq.(\ref{eq20}) and Eq.(\ref{eq21}) for 2-D and 3-D cases with ensemble size $10$, respectively, the black line and the orange line represent the results of thermodynamic integration method with ensemble size ${{10}^{9}}$.}
    \end{figure}

    Similarly, the numerical results of 2-D and 3-D case are presented in Fig.(\ref{fig5}), Fig.(\ref{fig6}) and Fig.(\ref{fig7}). The Fig.(\ref{fig5}) and Fig.(\ref{fig6}) shows the curve of FED changing with the particle density of terminal state when the temperature of the initial state and terminal state is fixed and the Fig.(\ref{fig7}) shows the curve of FED changing with the reverse temperature of terminal state when the volume of the initial state and terminal state is fixed. Here, we also set ${{F}_{a}}=0$ and redefine particle density $\rho \equiv N/V$. The periodic boundary condition is taken in these computations, the definitions of other parameters are the same as the 1-D case. The canonical ensemble Monte Carlo algorithm is used to generate the microscopic states in these figures. The definitions of the parameters in computation are the same as Eq.(\ref{eq20}) and Eq.(\ref{eq21}), the ensemble size is $10$ and the corresponded uncertainty of $\Delta f$ is smaller than $2\times {{10}^{-7}}$.The ensemble size of the thermodynamic integration method is ${{10}^{9}}$and the corresponded uncertainty of $\Delta f$ is smaller than $5\times {{10}^{-5}}$. The effectiveness of Eq.(\ref{eq20}) and Eq.(\ref{eq21}) can be confirmed because the data points of two methods meet well. Besides, the computational time of Eq.(\ref{eq20}) and Eq.(\ref{eq21}) is far less than the thermodynamic integration method, which suggests the advantages of the equilibrium equality for FED are more apparent in multiple dimensional models.

    \section{Summary}\label{sec6}
    The equilibrium equality proposed by this article has expressed FED between any two equilibrium states as the average over one canonical ensemble successfully, which means that the calculation of FED between any two states can be achieved by applying the parameters of two states and generating one canonical ensemble. In comparison, the JE method needs to calculate the work done in a non-equilibrium process and the thermodynamic integration method relies on the path, the advantage of the equilibrium equality is obvious that it makes the "skip" between equilibrium states practical. The calculation based on the equilibrium equality needs less samples when it reaches the same accuracy as JE method or thermodynamic integration method, which promote the efficiency of computation. It is meaningful to researches on systems with large size or the computation of FED with high accuracy and its efficiency is examined by the related numerical study. Besides, the equality reduces the calculation of FED to a sampling problem in a position vector space so the calculation can be completed by applying the Monte Carlo algorithm directly and it can be developed by some enhanced sampling technique, which makes the structure of the program more simple and universal. What's more, this equality completes the full generalization of the FEPT and thus make the theories for FED calculation more completed.

\bibliography{ref}
\end{document}